\documentclass[12pt]{iopart}
\usepackage{graphicx}
\usepackage{iopams}
\usepackage{amsthm,amsfonts,amssymb}

\begin{document}
\title[In situ comparison of the critical current density in {\rm YBCO} thin films]{In situ comparison of the critical current density in $\mathrm{Y}\mathrm{Ba}_2\mathrm{Cu}_3\mathrm{O}_{7-\delta}$ thin films measured by the screening technique under two criteria}

\author{F. Gamboa$^{1}$, V\'ictor Sosa$^1$, I. P\'erez$^2$, J.A. Matutes-Aquino$^3$ and A. Moewes$^2$}
\address{$^1$ Departamento de F\'isica Aplicada,
Cinvestav Unidad M\'erida, Km 6 Ant. Carretera a Progreso, A.P. 73, C.P. 97310. M\'erida, Yucat\'an M\'exico}
\address{$^2$ Physics Department and Engineering Physics,  University of Saskatchewan, 116 Science Place, S7N 5E2, Saskatoon, SK Canada}
\address{$^3$ Departamento de F\'isica de Materiales, Centro de Investigaci\'on en Materiales Avanzados, S.C. (CIMAV) Ave. Miguel de Cervantes 120, Complejo Industrial Chihuahua, Chihuahua, M\'exico}
\ead{ffgamboa@mda.cinvestav.mx}

\begin{abstract}
\label{abstract}
In this investigation we report the determination of the critical current density $J_c$  flowing in a $\mathrm{Y}\mathrm{Ba}_2\mathrm{Cu}_3\mathrm{O}_{7-\delta}$ superconducting thin film. Estimation of $J_c$ was carried out by an inductive technique, the so-called screening technique, in which both the imaginary part of the fundamental harmonic of susceptibility $\chi''_1$ and third harmonic voltage $V_3$ criteria were considered for the determination of the full penetration field.
 In order to verify the reliability of this technique under two criteria, we investigated the homogeneity of $J_c$ via transport measurements conducted on four microbridges patterned in the same film. Based on the transport method we found that both techniques yield similar results in the determination of $J_c$ for temperatures close to critical temperature $T_c$. However, at temperatures relatively far from $T_c$, the $V_3$ criterion showed a better agreement with the transport data. Furthermore, using both criteria, we propose a methodology to estimate in situ the coil factor associated with the $V_3$ criterion, avoiding in this way the need of implementing an additional technique.
\end{abstract}

\section{Introduction}
\label{introduction}
The application of high-temperature superconductors (HTSC) in the electronics industry is a promising enterprise, which goes from SQUID{'}s to integrated circuits to superconducting computers. Therefore, it is of prime importance to understand the superconducting properties of these materials not only of bulk samples but of thin films as well. In recent years, a great deal of attention has been put in studying properties of superconducting thin films. To determine the superconducting properties such as the critical current density $J_{c}$ of either a bulk or a thin film superconductor, researches often use the traditional four-probe technique. Although this technique is quite satisfactory for bulk samples, it is highly intrusive for thin films and, as a consequence, most of the times they can no longer be used in further tests. To overcome this problem, several non-invasive techniques have been proposed. So far, two inductive techniques of this kind are widely used. In one of these cases known as the conventional ac susceptibility technique\cite{nikolo,goldfarb}, the sample is embedded in a pick-up coil, which in turn is coaxially mounted inside a driving coil. The experimental arrangement is sensitive to the overall magnetic response of the sample when it is subjected to the influence of an ac magnetic field of the form $H(t)$ = $H_{0} \cos{(\omega t)}$, 
where $H_{0}$ is the amplitude, $\omega$ the frequency and $t$ the time. The induced voltage in the pick-up coil is zero when the sample is in its normal state but as the superconducting transition occurs an unbalanced voltage sets in \cite{nikolo}. Due to the nonlinear magnetization of the sample, harmonics of the voltage will emerge. It has been shown that harmonics of the magnetic susceptibility are proportional to the harmonics of the voltage. With the help of this technique and based on analysis of the imaginary part of the fundamental complex magnetic susceptibility as a function of the temperature $\chi''_{1}(T)$, Xing \etal \cite{xing} proposed a method to determine the $J_{c}$ in superconducting thin films.\\ 
The other alternative, the mutual inductive technique or screening technique\cite{claassen,mawatari1}, offers a practical approach not only to determine the $J_c$\cite{mawatari1,yamasaki2,yamasaki,yamada} but also to measure the harmonics of the magnetic susceptibility\cite{israel1,israel2}. In this experimental setup the film is placed between two small coils forming a sandwich arrangement, and as in the conventional ac susceptibility technique, the induced voltage is measured through the pick-up coil. The reliability of the technique can be guaranteed if the size of the sample is at least twice larger than the outer radius of the coils \cite{claassen}. Previous experiments\cite{yamasaki2,yamasaki,yamada} have shown that the amplitude of the third harmonic voltage $V_3$ emerges when the amplitude $I_0$ of the driving current reaches a threshold current $I_{th}$ which can be related to the $J_c$ of the film.\\
Therefore, there are two criteria to estimate the $J_c$ independent of the technique used. For brevity we will call them: the $\chi''_{1}(T)$ criterion and the $V_{3}(I_0)$ criterion. Although both criteria are widely used there is not consensus in their convergence and no fundamental theory to single out one or the other. \\
In this investigation, we realized measurements of $J_c$ in $\mathrm{Y}\mathrm{Ba}_2\mathrm{Cu}_3\mathrm{O}_{7-\delta}$ superconducting thin films using both criteria as measured with the screening technique only. Then, we compare our results with the conventional four-probe method.
We found that both criteria yield similar results for the temperatures close to $T_c$, however for temperatures relatively far from $T_c$ the corresponding curves begin to diverge. Moreover, the analysis of these criteria sheds light on the determination of the coil factor $k$ from the $V_{3}(I_0)$ criterion and therefore, its dependence on other experimental techniques is not necessary.
\section{Theory}
\label{theory}
The experimental geometry used in this work is shown in Fig. 1. An ac magnetic field $H$ = $H_{0}\cos{(\omega t)}$ has been applied along the c axis of the film. The field is generated just above of the film by a current  $I$ = $I_{0}\cos{(\omega t)}$ passing through the drive coil. Then, the superconducting response is registered through the pick-up coil placed on the rear of the film. The induced voltage depends on the coupling between the coils and is strongly influenced by the superconducting behavior of the film. 
\subsection{The $\chi_1''(T)$ criterion}
\label{theory_1}
Xing et al.\cite{xing} used the imaginary part of fundamental harmonic of susceptibility $\chi''_{1}(T)$ to estimate $J_c$. He found that the magnetic moment or the shielding current in the hysteresis cycle saturates when the amplitude $H_{0}$ reaches the penetration field $H^{*}$, which is related to $J_c$ by 
\newline 
\begin{equation}
\label{eq1}
H^*= \frac{J_{c}d}{3}
\end{equation}
where $d$ is the film thickness. Based on the Sun model\cite{sun} the imaginary part of the fundamental harmonic for a thin film is found to be
\newline
\begin{equation}\label{eq2}
\chi''_{1}\propto\frac{e^{-h}}{h}[h\cosh{(h)}-\sinh{(h)}]
\end{equation}
where $h = H_{0}/H^*$ is the reduced field. A plot of $\chi''_{1}$ versus $h$ shows a peak at $h =1.344$ or $H_{0}$ = 1.344$H^*$. If  $\chi''_{1} $ is plotted as a function of the temperature, a peak appears at a temperature $T_p$ where the above relation is satisfied, i.e., $H_0$ = 1.344$H^{*}(T_p)$. 
From the last relation and \Eref{eq1}, the following formula is obtained\cite{xing}:\\
\begin{equation}
\label{eq3}
J_{c}(T_p)= 3.157 \frac{H_{0}}{d}.
\end{equation}
Here $d$ is given in meters and $H_{0}$ is the rms value in A/m. Since $V''_{1}(T)$ $\propto\omega H_{0}\chi''_1$  \cite{israel1} one can obtain $J_c$ by plotting $V''_{1}(T)$ as a function of the temperature for distinct magnetic field amplitudes. So, for each curve a peak appears at a temperature $T_p$, and using the previous relation (\ref{eq3}), we can obtain the temperature dependence of $J_c$.
\subsection{The $V_{3}(I_{0})$ criterion}
\label{theory_2}
Consider that the sample is held at a constant temperature $T$ ($T<T_c$) and a small field $H$ has been applied to the film. If the value of $I_0$ is smaller than a certain threshold value $I_{th}$ ($I_0<I_{th}$), the magnetic field under the film is screened by a superficial superconducting current $K_s$ (sheet current) that flows in the film. The magnetic field amplitudes at the upper and lower surfaces of the film are, $H_1$ = 2$H_0$ and $H_2$ = 0, respectively\cite{claassen,mawatari1}. In this case, the film is regarded as an image coil reflected through the upper surface of the film, carrying the same current but in the opposite direction. The magnetic response of the film is linear, and no harmonics of the voltage are induced in the pick-up coil. When $I_{0}$ = $I_{th}$, the magnetic field achieves the full penetration and the film response is no longer lineal. As a result, the third-harmonic voltage $V_3$ in the pick-up coil starts to emerge. At this point, the amplitude of the ac magnetic field near the surface of the film is $H_1$ = 2$H_0$ = $J_{c}d$\cite{claassen,mawatari1}. Then, $I_{th}$ can be expressed as $I_{th}$ = ($d/k$)$J_c$, where $k$ is a coil factor and $J_c$ can be determined by measuring the occurrence of $V_3$ with increasing $I_0$. The coil factor depends on the shape and the location of the experimental coils. Then, to determine its value it is necessary to measure the $J_c$ by implementing an additional technique. 
\begin{figure}[!tbp]
\begin{center}
\includegraphics[scale=0.7]{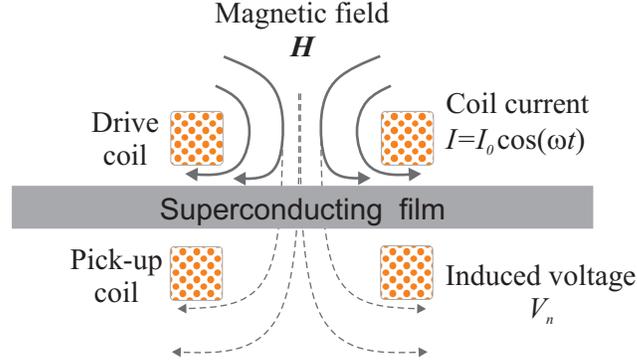}
\caption{The screening technique: transversal section. A pair of small and coaxial coils with the film of interest interposed between them. The magnetic field is generated by the driving current $I_0 \cos{\omega t}$ that flows through of the drive coil mounted just above a superconducting thin film. The pick-up coil detects the magnetic response at the rear region of the film (dotted lines represent the lines of the magnetic field when the film is in the normal state).}
\label{twocoils}
\end{center}
\end{figure}
\section{Experimental}
\label{experimental}
\begin{figure}[!tbp]
\begin{center}
\includegraphics[scale=0.8]{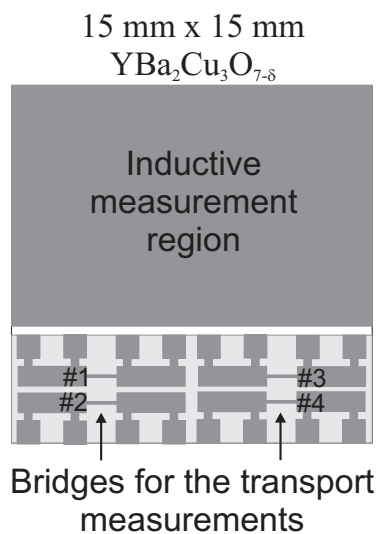}
\caption{The 15 mm x 15 mm $\mathrm{Y}\mathrm{Ba}_2\mathrm{Cu}_3\mathrm{O}_{7-\delta}$ thin film used in this work (Theva Co.). Upper zone for inductive measurements and lower zone with four patterned bridges (etched)}
\label{cal_film}
\end{center}
\end{figure}

Figure \ref{cal_film} shows the measurements regions patterned on the 15 mm $\times$ 15 mm $\mathrm{Y}\mathrm{Ba}_2\mathrm{Cu}_3\mathrm{O}_{7-\delta}$ thin film (Theva Co.) used in this work. The film of thickness $d$ = 200 nm, was deposited on a $\mathrm{Ce}\mathrm{O}_2$-buffered R-cut $\mathrm{Al}_{2}\mathrm{O}_3$ substrate by the so-called thermal co-evaporation method and was coated in situ with a 200 nm layer of Au. The growth and patterned details were reported elsewhere\cite{irgmaier}. For comparison and with the purpose of evaluating  the precision of both criteria, $J_c $ was also determined from transport measurements by the conventional four-probe method. The film has two measuring sections. A section for inductive measurements and a section for transport measurements where four microbridges of 1 mm length and 40 $\mu $m width can be found. In the inductive region the pair of coils were placed at the center, as scketched in the Fig. \ref{twocoils}. Each coil had 360 turns, an outer diameter of 5 mm, inner diameter of 1 mm, and height of 1 mm.\\
\begin{figure}[!tbp]
\begin{center}
\includegraphics[scale=1]{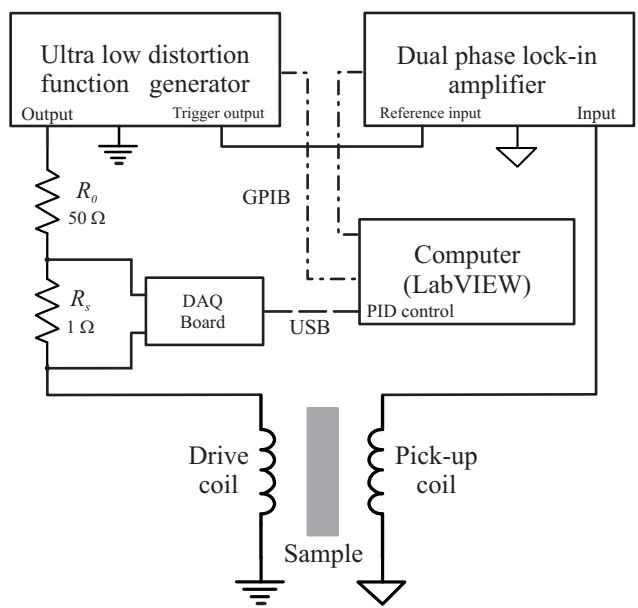}
\caption{Block diagram of the experimental setup for the inductive measurements.}
\label{sist_screening}
\end{center}
\end{figure}
The block diagram of the experimental setup for inductive measurements is shown in Fig. \ref{sist_screening}. The sinusoidal voltage was supplied from a low distortion generator (SRS-DS360) to the drive coil. For the $V_3$ criterion, the current was determined measuring the drop voltage in the resistor $R_{s}$ in series with the drive coil by means of a data acquisition board (NI-USB6216). For the case of the $\chi''_{1}(T)$ criterion, we wish to maintain a constant magnetic field as the temperature is changed, so the current in $R_s$ was controlled by means of a Proportional-Integral-Derivative (PID) control implemented in a computer program. In both criteria, a phase-sensitive lock-in amplifier (SRS-SR830) was used to analyze the signal coming from the pick-up coil. The superconducting sample was cooled down by means of a He cold finger coupled to a cryostat (Cryotech-ST15) compressor and the temperature was controlled via a temperature controller (SI-9600) and monitored by a diode (SI-410A) placed near the sample.
\section{Experimental results and discussion}
\label{results}
\begin{figure}[!tbp]
\centering
\includegraphics[scale=1]{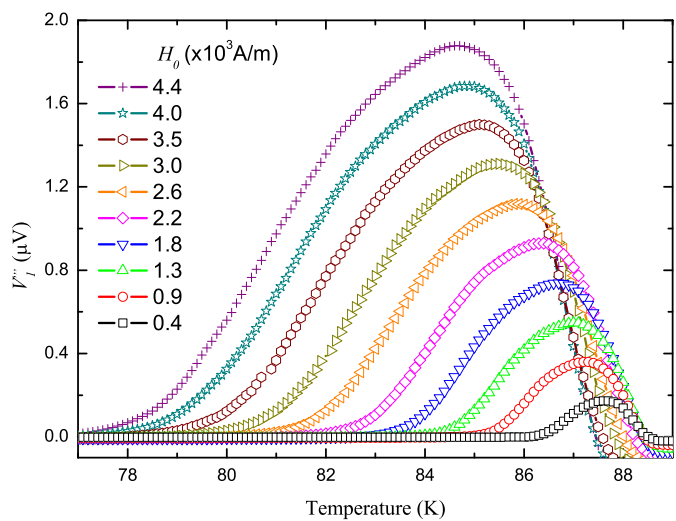}
\caption{Temperature dependence of the $V''_1$ for several magnetic field amplitudes $H_0$. Both $V''_1$ and $H_0$ are given in rms values.}
\label{v1_T}
\end{figure}

\begin{figure}[!tbp]
\centering
\includegraphics[scale=1]{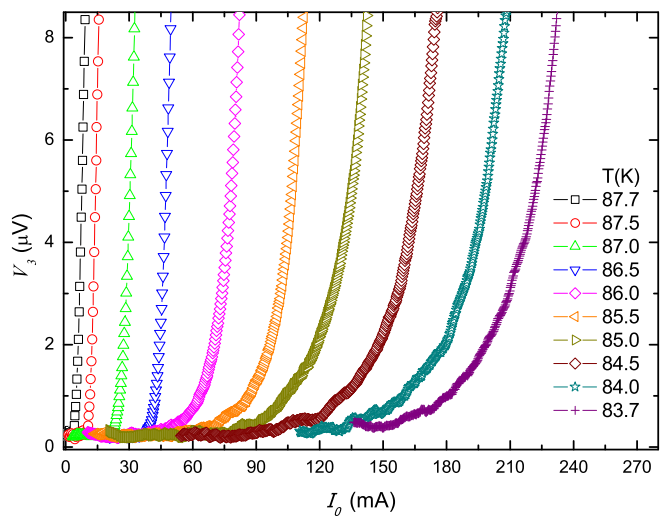}
\caption{Driving current dependence of $V_3$ for several temperatures. Both $I_0$ and $V_3$ are given in rms values.}
\label{v3_I}
\end{figure}

\begin{figure}[!tbp]
\centering
\includegraphics[scale=1]{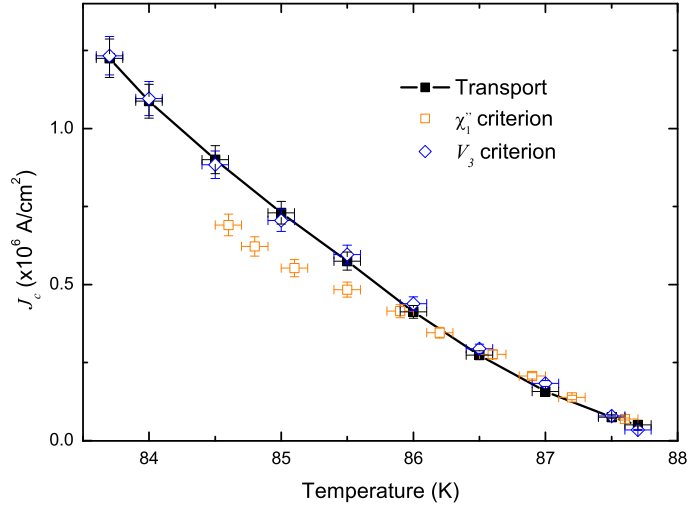}
\caption{Dependence of the $J_c$ obtained by the transport technique and the screening technique for temperatures varying from 83.7 to 87.7 K. For the screening technique, the $J_c$ under the $V_3$ and $\chi''_1$ criteria are presented.}
\label{Jc_todos}
\end{figure}

We first characterized the bridges as of $J_c$ and $T_c$. Table \ref{proper} shows a summary of the results for the four bridges. The transport measurements were carried out in a liquid nitrogen bath with the conventional four-probe method and a criterion of electric field $E$ = 1$\times 10^{-3}$ V/m was applied. We could verify that both $T_c$ and $J_c$ for the four bridges are quite homogeneous along the sample.\\
\begin{table}[!tbp]
\caption{$T_c$ and $J_c$ measurements for the bridges at positions $\#1-\#4$. $J_c$ values defined at $E$ = 1$\times 10^{-3}$ V/m.}
\centering 
\begin{tabular}{c|cl}
\hline
\footnotesize{Bridge} & \footnotesize{$T_c$ (\textrm{K}) \textrm{$\pm$ 0.1 K}} & \footnotesize{$J_c$ ($\times10^{6}$\textrm{A/cm$^2$}) at 77.3 \textrm{K}}
\\ \hline\hline
\footnotesize{$\#1$} & \footnotesize{88.0} & \footnotesize{3.545 ($-0.12\%$)}\\ 
\footnotesize{$\#2$} & \footnotesize{88.0} & \footnotesize{3.549 ($+1.31\%$)}\\ 
\footnotesize{$\#3$} & \footnotesize{88.0} & \footnotesize{3.547 ($+0.12\%$)}\\
\footnotesize{$\#4$} & \footnotesize{88.0} & \footnotesize{3.548 ($-1.28\%$)}\\ 
\footnotesize{$\mathrm{Average}$} & \footnotesize{88.0} & \footnotesize{3.542}\\ 
\hline
\end{tabular}
\label{proper}
\end{table}
The next step was to estimate $J_c$ by the screening method according to the two criteria outlined above. The curves of the 
imaginary part of fundamental harmonic $V''_1$ as a function of the temperature for several magnetic field amplitudes $H_0$, are shown in the Fig. \ref{v1_T}.  Amplitudes of ac magnetic field ranging from 0.4 to 4.4$\times10^3$ A/m were applied to the film at a cooling rate of 2 K/min. From these curves we obtained the value of the temperature at the maximum ($T_p$) of each curve and use \Eref{eq3} to estimate $J_c(T_p)$ for a temperature range between 84.6 K and 87.6 K. We can see that as the amplitude increases the peak shifts towards lower temperatures and at the same time, the width of the curve increases. It is well known that these effects are due to the field and temperature dependence of $J_c$\cite{xing,israel2,sun}.
Then, we estimated $J_c$ by the $V_3$ criterion for a similar temperature range. The curves for the third harmonic of the voltage $V_3$ as a function of the driving current amplitude $I_0$ are shown in the Fig. \ref{v3_I}. Note that as the temperature decreases the curve of $V_3$ rises smoother than at temperatures close to $T_c$. From these results, the $I_{th}(T)$ curve was obtained. $I_{th}$ was estimated at the current $I_0$ when $V_3$ reached 1 $\mu$V. In order to obtain the $J_{c}(T)$ with this criterion, the coil factor $k$ was determined by fitting the $I_{th}(T)$ data with the $J_{c}$ curve obtained by means of the transport measurements. Then, a coil factor $k$ =139$\pm$23 $\textrm{cm}^{-1}$ was obtained.
The temperature dependence of $J_c$ for the four bridges obtained by transport and inductive measurements is shown in Fig. \ref{Jc_todos}, for temperatures ranging from 83.7 K to 87.7 K. The values of the transport measurements reported represent an average of the data corresponding to the different values obtained for each bridge. These curves are in good agreement among the different approaches.
However, if the $I_{th}(T)$ is fitted with the curve of $J_c$ from $\chi''_1$ criterion, then we get $k$ = 127$\pm$19 $\textrm{cm}^{-1}$, the same value found previously.
Convergency of both inductive criteria near $T_c$ have been previously reported by Acosta \etal \cite{acosta}. However, their measurements reached a wider temperature range (possibly due to a lower $J_c$ of their sample), but their resolution near $T_c$ was poor and therefore a useful comparison with our results was not feasible. In addition, they determined the $J_c$ with the $V_3$ approach but taking the value of the driving current as the voltage starts to emerge and applying the \Eref{eq1}.  For the $J_c$ curves based on the $\chi''_1$ criterion they determined the $H^*$ considering the value of the temperature at the peak of the curve of the fundamental harmonic of voltage. Experimentally we found that both voltages start to emerge at the same amplitude of the magnetic field (proportional to the drive coil current), leading to an overestimation of $J_c$ according to the $\chi''_1$ criterion, as finally they conclude.\\
On the other hand, we noted that for lower temperatures, deviations from the $\chi''_1$ criterion are observed. Because the determination of $J_c$ depends on the specific selection of the full penetration field $H^*$, then we are observing the behavior of the $H^*$ from each criterion. It follows that both criteria converge near $T_c$ and they start to deviate as the temperature moves away from $T_c$. Similar behavior of the fields has been reported by Shaulov\cite{shaulov} in sintered YBCO samples that were studied with conventional ac susceptibility technique.
\section{Summary}
\label{summary}
In this research the $V_3$ and $\chi''_1$ inductive approaches were confronted with the conventional transport method. We found that both criteria yield similar results for temperatures close to $T_c$ and agree with that transport measurements. However, for lower temperatures the $J_c$ curves based on the $\chi''_1$ criterion diverge considerably whereas the $V_3$ criterion agree pretty well along the whole temperature treated in this work.\\
Our investigation also provided a methodology to estimate in situ the $k$ coil factor associated with the $V_3$ criterion.
This is achieved by using the $\chi''_1$ criterion without the need of any additional technique. 
\section*{Acknowledgements}
F. Gamboa wishes to acknowledge a scholarship from CONACYT and I. P\'erez also wishes to acknowledge the financial support from CONACYT under grant 186142. The authors wish to thank  O. G\'omez for the technical support.


\section*{References}
\begin{thebibliography}{99}

\bibitem{nikolo} M. Nikolo, ``Superconductivity: A guide to alternating current susceptibility measurements and alternating current susceptometer design," Am. J. Phys., vol. 63, no. 1, pp. 57-65, 1995.

\bibitem{goldfarb} R. Goldfarb, M. Lelental, and C. Thompson, ``Magnetic susceptibility of superconductors and
other spin systems. Plenum Press, 1991, ch. Alternting-field susceptometry and magnetic susceptibility of superconductors," pp. 49-80.

\bibitem{xing} W. Xing, B. Heinrich, J. Chrzanowski, J. Irwin, H. Zhou, A. Cragg, and A. Fife, ``Determination
of critical current densities of $\mathrm{Y}\mathrm{Ba}_2\mathrm{Cu}_3\mathrm{O}_{7-\delta}$ thin films from ac susceptibility measurements", Physica C, vol. 205, pp. 311-322, 1993.

\bibitem{claassen} J. Claassen, M. Reeves, and R. Soulen., ``A contactless method for measurement of the critical
current density and critical temperature of superconducting films," Rev. Sci. Instrum., vol. 62, no. 4, pp. 996-1004, 1991.
\bibitem{mawatari1} Y. Mawatari, H. Yamasaki, and Y. Nakagawa, ``Critical current density and third-harmonic voltage in superconducting films," Appl. Phys. Lett., vol. 81, no. 13, pp. 2424-2426, 2002.

\bibitem{yamasaki} H. Yamasaki, Y. Mawatari, Y. Nakagawa, and H. Yamada, ``Nondestructive, inductive measurement of critical current densities of superconducting films in magnetic fields," IEEE
Trans. Appl. Supercond., vol. 13, no. 2, pp. 3718-3721, 2003.

\bibitem{yamasaki2} H. Yamasaki, Y. Mawatari, and Y. Nakagawa, ``Precise determination of the threshold current for third-harmonic voltage generation in the ac inductive measurement of critical current densities of superconducting thin films," IEEE Trans. Appl. Supercond., vol. 15, no. 2, pp. 3636-3639, 2005.

\bibitem{yamada} H. Yamada, T. Minakuchi, D. Itoh, T. Yamamoto, S. Nakagawa, K.Kanayama, K. Hirachi, Y. Mawatari, and H. Yamasaki, ``Variable RL-cancel circuit for precise $J_c$ measurement using third-harmonic voltage method," Physica C, vol. 451, pp. 107-112, 2007.

\bibitem{israel1}  I. O. P\'erez-L\'opez, F. Gamboa, and V. Sosa, Critical current density and ac harmonic voltage
generation in $\mathrm{Y}\mathrm{Ba}_2\mathrm{Cu}_3\mathrm{O}_{7-\delta}$ thin films by the screening technique", Physica C, vol. 470, pp. S972-S974, 2010.

\bibitem{israel2} I. P\'erez, V. Sosa, and F. Gamboa, ``Study of high-harmonics of complex ac susceptibilty in
$\mathrm{Y}\mathrm{Ba}_2\mathrm{Cu}_3\mathrm{O}_{7-\delta}$ thin films by the mutual inductive method", Physica C, vol. 470, pp. 2061-2066, 2010. 

\bibitem{sun} J. Sun, M. Scharen, L. Bourne, and J. Schrieer, ``ac susceptibility of the thin-film superconductors
near Tc: A theoretical study," Phys. Rev. B, vol. 44, no. 10, pp. 5275-5279, 1991.

\bibitem{irgmaier} K. Irgmaier, R. Semerad, W. Prusseit, A. Ludsteck, G. Sigl, H. Kinder, J. Dzick, S. Sievers,
H. Freyhardt, and K. Peters, ``Deposition and microwave performance of YBCO films on technical ceramics," Physica C, vol. 372-376, p. 554, 2002.

\bibitem{acosta} C. Acosta, M. Acosta, V. Sosa, and O. Ares, ``Comparative analisys of the determination of the
$J_c$ of YBCO films at different temperatures and magnetic fields by means of the shielding technique," Physica C, vol. 341-348, pp. 2051-2052, 2000.

\bibitem{shaulov} A. Shaulov and D. Dorman, ``Investigation of harmonics generation in alternating magnetic
response of superconducting YBCO," Appl. Phys. Lett., vol. 53, no. 26, pp. 2680-2682, 1988.

\end{thebibliography}
\end{document}